\documentclass[aps,prc,twocolumn,superscriptaddress,showpacs,showkeys,amsmath,amssymb,floatfix]{revtex4}
\usepackage{booktabs,graphicx,mathrsfs,verbatim}
\usepackage[usenames,dvipsnames]{color}
\usepackage{multirow,bigdelim}
\usepackage{ulem}
\usepackage{color}
\usepackage[sort&compress]{natbib} 
\newcommand{\valencia}{\affiliation{Instituto de F{\'i}sica Corpuscular, CSIC-Universidad de Valencia, E-46071 Valencia, Spain}}
\newcommand{\osakadep}{\affiliation{Department of Physics, Osaka University, Toyonaka, Osaka 560-0043, Japan}}
\newcommand{\bordeaux}{\affiliation{Centre d'Etudes Nucl{\'e}aires de Bordeaux Gradignan, CNRS/IN2P3 - Universit{\'e} de Bordeaux, 33175 Gradignan Cedex, France}}
\newcommand{\surrey}{\affiliation{Department of Physics, University of Surrey, Guildford GU2 7XH, Surrey, UK}}
\newcommand{\debrecen}{\affiliation{Inst. of Nuclear Research of the Hung. Acad. of Sciences, Debrecen, H-4026, Hungary}}
\newcommand{\instanbul}{\affiliation{Department of Physics, Istanbul University, Istanbul, 34134, Turkey}}
\newcommand{\caen}{\affiliation{Grand Acc{\'e}l{\'e}rateur National d'Ions Lourds (GANIL), CEA/DRF-CNRS/IN2P3, Bvd Henri Becquerel, 14076 Caen, France}}
\newcommand{\osakarcnp}{\affiliation{Research Center for Nuclear Physics, Osaka University, Ibaraki, Osaka 567-0047, Japan}}
\newcommand{\chile}{\affiliation{Comisi{\'o}n Chilena de Energ{\'i}a Nuclear, Casilla 188-D, Santiago, Chile}}
\newcommand{\belgium}{\affiliation{SCK.CEN, Boeretang 200, 2400 Mol, Belgium}}
\newcommand{\argonne}{\affiliation{Physics Division, Argonne National Laboratory, Argonne, Illinois 60439, USA}}

\begin{document} 

\title{Observation of the 2$^+$ isomer in $^{52}$Co}

\author{S.~E.~A.~Orrigo}\email{sonja.orrigo@ific.uv.es}\valencia
\author{B.~Rubio}\valencia
\author{W.~Gelletly}\valencia\surrey
\author{B.~Blank}\bordeaux
\author{Y.~Fujita}\osakadep\osakarcnp
\author{J.~Giovinazzo}\bordeaux
\author{J.~Agramunt}\valencia
\author{A.~Algora}\valencia\debrecen
\author{P.~Ascher}\bordeaux
\author{B.~Bilgier}\instanbul
\author{L.~C{\'a}ceres}\caen
\author{R.~B.~Cakirli}\instanbul
\author{G.~de~France}\caen
\author{E.~Ganio{\u{g}}lu}\instanbul
\author{M.~Gerbaux}\bordeaux
\author{S.~Gr{\'e}vy}\bordeaux
\author{O.~Kamalou}\caen
\author{H.~C.~Kozer}\instanbul
\author{L.~Kucuk}\instanbul
\author{T.~Kurtukian-Nieto}\bordeaux
\author{F.~Molina}\valencia\chile
\author{L.~Popescu}\belgium
\author{A.~M.~Rogers}\argonne
\author{G.~Susoy}\instanbul
\author{C.~Stodel}\caen
\author{T.~Suzuki}\osakarcnp
\author{A.~Tamii}\osakarcnp
\author{J.~C.~Thomas}\caen
%

\begin{abstract}
We report the first observation of the 2$^+$ isomer in $^{52}$Co, produced in the $\beta$ decay of the 0$^+$, $^{52}$Ni ground state. We have observed three $\gamma$-rays at 849, 1910, and 5185 keV characterizing the $\beta$ de-excitation of the isomer. We have measured a half-life of 102(6) ms for the isomeric state. The Fermi and Gamow-Teller transition strengths for the $\beta$ decay of $^{52m}$Co to $^{52}$Fe have been determined. We also add new information on the $\beta$ decay of the 6$^+$, $^{52}$Co ground state, for which we have measured a half-life of 112(3) ms.
\end{abstract}



\maketitle

\section{\label{intro}Introduction}

$\beta$-decay spectroscopy is a fundamental tool for the investigation of the nuclear structure of unstable nuclei \cite{Bohr1969,Blank2008,Rubio2009,PhysRevLett.112.222501}. Many neutron-deficient $fp$-shell nuclei lie on the astrophysical $rp$-process reaction pathway. Accordingly the study of the $\beta$ decay of such nuclei is of importance because it provides input to calculations of the $rp$-process and models of X-ray bursters \cite{Schatz1998167,Parikh2013225}. The investigation of the decay of these nuclei is difficult because they lie far away from stability. The odd-odd nuclei are particularly difficult to study because there are often two long-lived states, one of which is the ground state, with similar half-lives. This makes it hard to disentangle the two decays. This is because both states are in general members of the same two-particle multiplet and have therefore very similar structure. The only difference is how the spins of the valence nucleons couple to make the final spin. One strong contribution to the half-life is given by the Fermi transition, which is very fast and has identical strength in the two cases. How different the total half-life will be for the two states will thus depend on the details of the Gamow-Teller transitions. Here we present information on one such case. $^{52}$Co is a $T_z = -1$ odd-odd isotope in the $f_{7/2}$ shell that was first observed in an experiment performed at GANIL \cite{Pougheon1987}. There had been previous indications of the existence of a long-lived $\beta$-decaying excited state but it had not been isolated experimentally \cite{Hagberg1997183}.

The $^{52}$Mn nucleus, the mirror of $^{52}$Co, has a 2$^+$ isomeric state at 378 keV above the 6$^+$ ground state. This $^{52}$Mn isomer, having a half-life of 21.1(2) min \cite{PhysRev.113.602,NDS2015}, decays via two branches, 98.22(5)\% by $\beta^+$ decay to $^{52}$Cr and 1.78(5)\% via an internal transition to the ground state \cite{PhysRevC.16.1581,NDS2015}. Assuming isospin symmetry, a 2$^+$ isomeric state is also expected in $^{52}$Co at a similar energy. This would mean that we have the case of two states with \mbox{$J^{\pi}$ = 2$^+$} and 6$^+$, corresponding to the 2$^+$ and 6$^+$ members of the $(\pi f_{7/2})^{-1}(\nu f_{7/2})^{-3}$ multiplet.  The Fermi partial half-life will be of the order of 200 ms, and the total half-life will depend on the distribution and population of the low-lying 1$^+$, 2$^+$, 3$^+$ states in the $^{52}$Fe daughter for the decay of the 2$^+$ isomer, and of the 5$^+$, 6$^+$, 7$^+$ states in $^{52}$Fe for the decay of the 6$^+$ ground state. For instance, in the very similar case of $^{44}$V, with probable structure $(\pi f_{7/2})^{3}(\nu f_{7/2})^{1}$, the two states with \mbox{$J^{\pi}$ = 6$^+$} and 2$^+$ have half-lives of 150(3) ms and 111(7) ms \cite{Hagberg1997183}, respectively.

We have studied the $\beta^+$ decay of $^{52}$Ni to $^{52}$Co in Ref. \cite{PhysRevC.93.044336}. A study of the high-spin states in $^{52}$Co has been carried out recently \cite{Bentley2016}. The $\beta^+$ decay of $^{52}$Co to $^{52}$Fe was studied in Ref. \cite{Hagberg1997183}. The $^{52}$Co ground state, having \mbox{$J^{\pi}$ = 6$^+$} and $T$ = 1, undergoes $\beta$ decay to its Isobaric Analogue State (IAS) in $^{52}$Fe at 5655 keV \cite{Hagberg1997183}. Since the proton separation energy is 7378(7) keV \cite{NDS2015} proton emission is not possible here. A cascade of four $\gamma$-rays (1329, 1942, 1535 and 849 keV) was reported in Ref. \cite{Hagberg1997183} corresponding to the de-excitation of the IAS in $^{52}$Fe through the sequence \mbox{6$^+ (T = 1) \rightarrow 6^+ \rightarrow 4^+ \rightarrow 2^+ \rightarrow 0^+$} [also shown in our decay scheme in Fig. \ref{decay}(b)]. The measured $\gamma$-ray intensities in Ref. \cite{Hagberg1997183} implied the existence of a 31(14)\% $\beta$ feeding to the first excited state in $^{52}$Fe at 849 keV (\mbox{$J^{\pi}$ = 2$^+$}), which is quite unlikely to be due to direct feeding from the 6$^+$ state considering the \mbox{$\Delta L$ = 4} difference between the parent and daughter states. It was therefore suggested in Ref. \cite{Hagberg1997183} that this anomaly could be explained by extra feeding associated with the decay of $^{52m}$Co, although no clear evidence could be found.

In the present paper we report the first observation of the 2$^+$ isomer in $^{52}$Co, which was populated in the $\beta$ decay of $^{52}$Ni. The trick here was not to look at $^{52}$Co as a direct product of the fragmentation reaction, but as a product of the decay of the 0$^+$, $^{52}$Ni ground state (see the partial decay scheme in Fig. \ref{decay}(a)]. This decay process directly populates the 0$^+ (T = 2)$ IAS in $^{52}$Co, which then de-excites via the sequence \mbox{0$^+ \rightarrow 1^+ \rightarrow 2^+$} \cite{PhysRevC.93.044336}. This is a much cleaner way to populate the expected 2$^+$ isomeric state. We have observed the $\gamma$-rays emitted following the $\beta$ decay of the isomer and measured its half-life. The $\beta$-decay Fermi and Gamow-Teller transition strengths, $B$(F) and $B$(GT), have been determined [an upper limit for $B$(F)]. Moreover, selecting the direct production of $^{52}$Co we could obtain data on the $\beta$ decay of the $^{52}$Co, 6$^+$ ground state, which allowed us to add new information on this decay and measure the half-life with improved precision.

\section{\label{exp}The experiment}

\begin{figure}[!b]
  \centering
	\includegraphics[width=1\columnwidth,trim={0 0.2cm 3.1cm 5.5cm},clip]{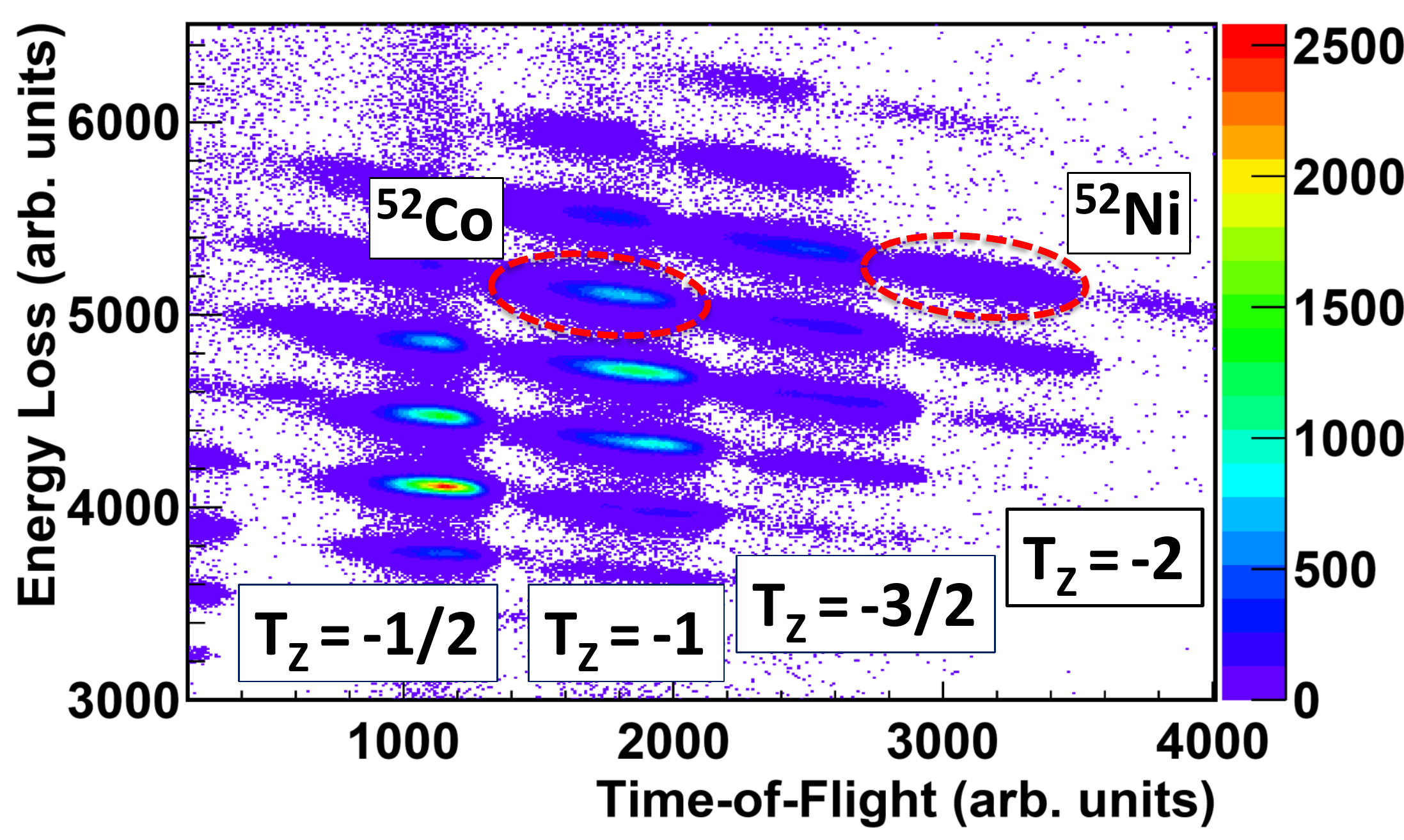}
	\vspace{-5.5 mm}
 	\caption{$\Delta E$ versus ToF identification plot for the dataset optimized to implant $^{56}$Zn close to the middle of the DSSSD (see Ref. \cite{PhysRevC.93.044336} for details). The positions of the $^{52}$Co and $^{52}$Ni implants are shown.}
  \label{ID1plot}
\end{figure}

\begin{figure}[!b]
  \centering
	\includegraphics[width=1\columnwidth,trim={0 0.2cm 3.0cm 5.6cm},clip]{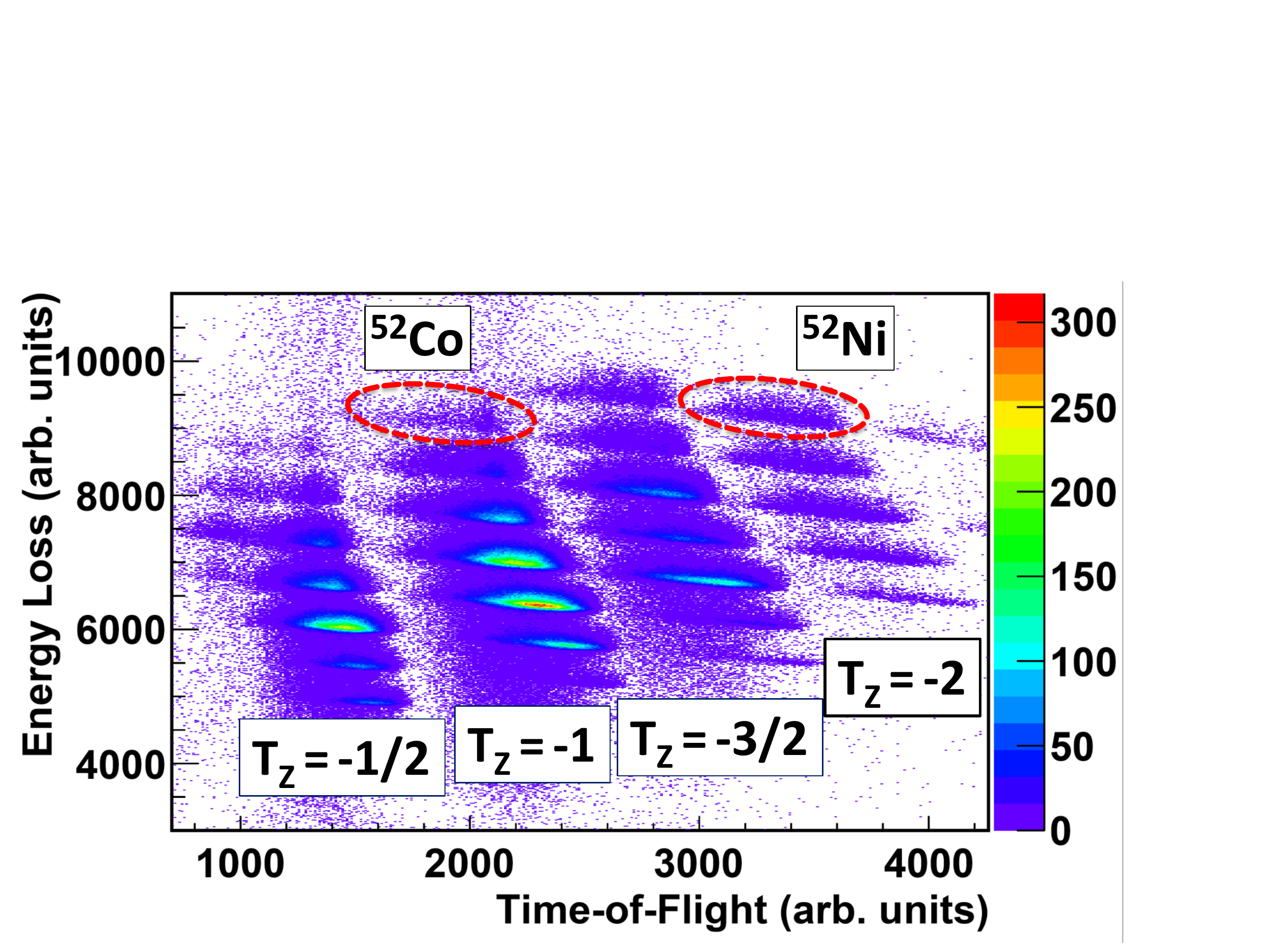}
	\vspace{-5.5 mm}
 	\caption{$\Delta E$ versus ToF identification plot for the dataset optimized for $^{48}$Fe. The positions of $^{52}$Co and $^{52}$Ni are shown.}
  \label{ID2plot}
  \vspace{-5.0 mm}
\end{figure}

We have studied the $\beta^+$ decay of $^{52}$Ni to $^{52}$Co in an experiment done at GANIL \cite{PhysRevC.93.044336}. $^{52}$Ni was produced by fragmenting a $^{58}$Ni$^{26+}$ primary beam, accelerated to 74.5 MeV/nucleon, on a 200-$\mu$m-thick natural Ni target. $^{52}$Co was also produced directly in the same experiment. After selection of the fragments in the LISE3 separator \cite{Anne1992276}, they were implanted into a 300-$\mu$m-thick Double-Sided Silicon Strip Detector (DSSSD). The DSSSD was used to detect both the implanted heavy ions and subsequent charged-particle ($\beta$ particles and protons) decays. Four EXOGAM Germanium clovers \cite{exogam} surrounding the DSSSD were used to detect the $\beta$-delayed $\gamma$-rays.

The ions were identified by combining the energy loss signal generated in a silicon $\Delta E$ detector located 28 cm upstream from the DSSSD and the Time-of-Flight (ToF), defined as the time difference between the cyclotron radio-frequency and the $\Delta E$ signal (see Figs.  \ref{ID1plot} and \ref{ID2plot}). An implantation event was defined by simultaneous signals from both the $\Delta E$ detector and the DSSSD. A decay event was defined by a signal above threshold in the DSSSD and no coincident $\Delta E$ signal.

The experimental setup is described in detail in Ref. \cite{PhysRevC.93.044336}, as well as the data analysis procedures employed.

\section{\label{results}The $^{52}$Co, 2$^+$ isomer}

\begin{figure*}[!ht]
  \centering
  \includegraphics[height=0.89\textheight,trim={0 0.35cm 0 0.4cm},clip]{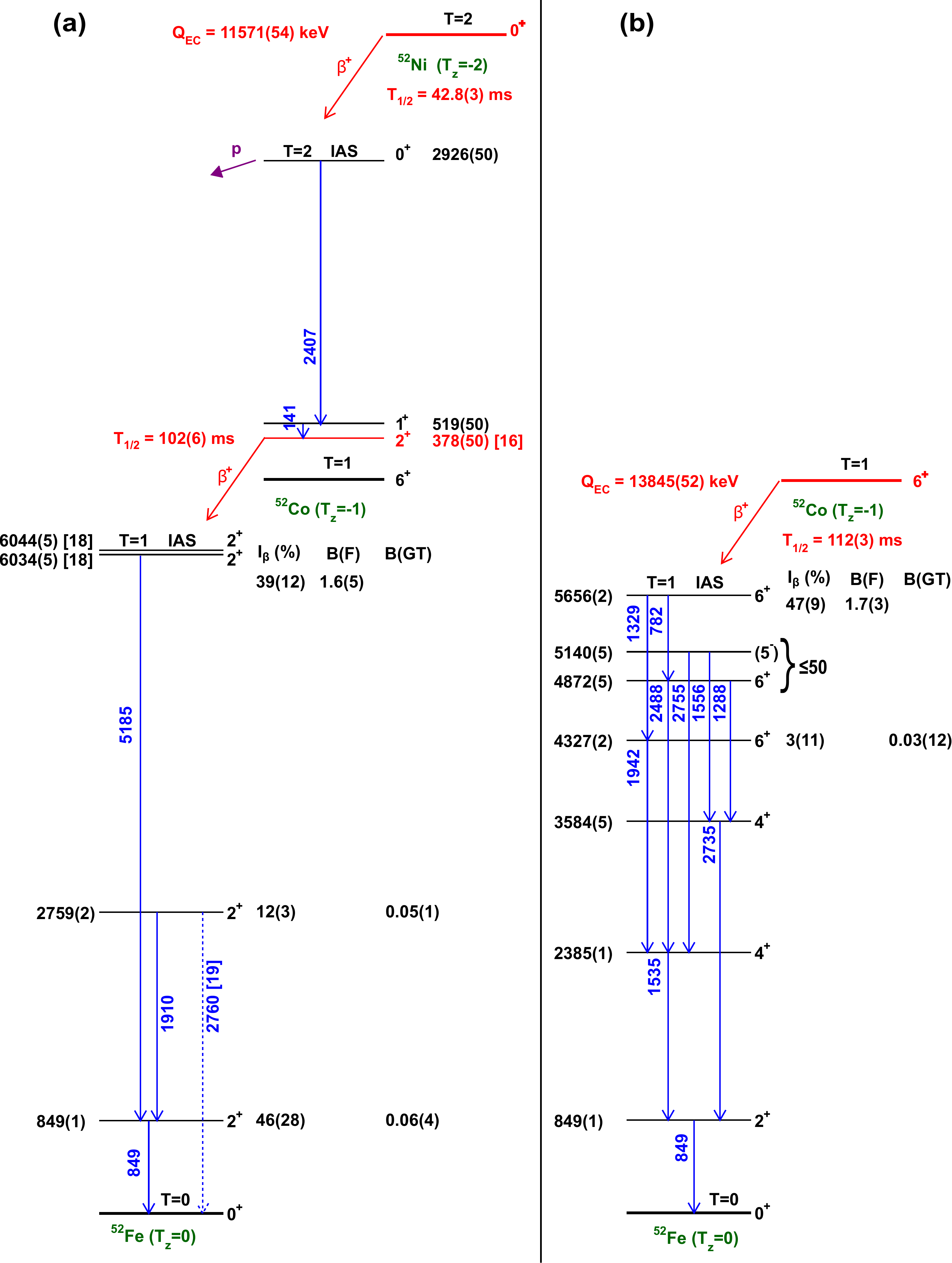}
	\vspace{-3.0 mm}
	\caption{(a) Partial decay scheme of $^{52}$Ni, including the decay of the $^{52}$Co, 2$^+$ isomer. Other 1$^+$ states populated in this decay and de-exciting via proton decay (see Ref. \cite{PhysRevC.93.044336}) are not included in the figure. The proton branching of the $^{52}$Co IAS is 25(5)\%. The energy of the isomeric level in $^{52}$Co, 378(50) keV, is assumed from the mirror $^{52}$Mn \cite{PhysRevC.7.677}. Two 2$^+$ levels separated by 10 keV are reported as IAS candidates in $^{52}$Fe \cite{DECOWSKI1978186}. The dashed $\gamma$-ray was reported in the literature \cite{JPSJ43} but not seen in the present work. The quoted $I_\beta$ branchings refer to 100 decays from $^{52m}$Co (2$^+$) estimated using the intensity of the 141 keV $\gamma$ line. (b) Decay scheme of the $^{52}$Co, 6$^+$ ground state deduced from the results of the present experiment. The quoted $I_\beta$ branchings refer to 100 decays from $^{52}$Co (6$^+$).}
	\label{decay}
\end{figure*}

In order to study the decay of the 2$^+$ isomer in $^{52}$Co, we have selected the events where $^{52}$Ni was implanted (see Figs.  \ref{ID1plot} and \ref{ID2plot}). For the following discussion we refer to the partial decay scheme shown in Fig. \ref{decay}(a), which starts from the $\beta^+$ decay of $^{52}$Ni to $^{52}$Co and then to $^{52}$Fe. The $\beta$ decay of $^{52}$Ni \cite{PhysRevC.93.044336} populates the 0$^+ (T = 2)$ IAS in $^{52}$Co at 2926(50) keV with a $\beta$ feeding of 56(10)\%, consistent with the expected Fermi strength \mbox{$B$(F) = $|N-Z|$ = 4}. Thereafter the decay of the IAS proceeds 25(5)\% of the time by proton emission to $^{51}$Fe and 75(23)\% of the time by a $\gamma$-ray cascade. The cascade consists of $\gamma$-rays of 2407 and 141 keV energy, with intensities $I_{\gamma}$ of 42(10)\% and 43(8)\%, respectively, and populating in sequence the levels at 519(50) \mbox{($J^{\pi}$ = 1$^+$)} and 378(50) \mbox{($J^{\pi}$ = 2$^+$)} keV in $^{52}$Co. As explained in detail in Ref. \cite{PhysRevC.93.044336}, we have assumed for the last level an energy of 378(50) keV from the value in the mirror nucleus $^{52}$Mn, 377.749(5) keV \cite{PhysRevC.7.677,NDS2015}, fixing in this way the excitation energies for the $^{52}$Co levels. The error of 50 keV on the 378 keV level energy, which accounts for possible mirror energy differences (MED), was estimated in Ref. \cite{PhysRevC.93.044336} by looking at the energies of the levels up to 400 keV in mirror nuclei with $T_z = +1/2, \text{-}1/2, +1, \text{-}1$. MED data for 2$^+$ states in the $A$ = 42-54 region \cite{Bentley2015} shows that our uncertainty is realistic and conservative. No $\gamma$-ray was observed from the 378 keV level, which is expected to be an isomeric state. This is not surprising since the $^{52}$Co, 2$^+$ level can decay by a Fermi $\beta$ transition, in contrast with its homologous state in the mirror nucleus. In $^{52}$Mn the Fermi transition is not possible and this, together with the smaller $Q_\beta$-value, makes the $\beta$ decay much slower and the slow E4 transition can compete with it.

\begin{figure}[!ht]
	\begin{minipage}{1.0\linewidth}
    \centering
		\vspace{-5.0 mm}
    \includegraphics[width=1\columnwidth]{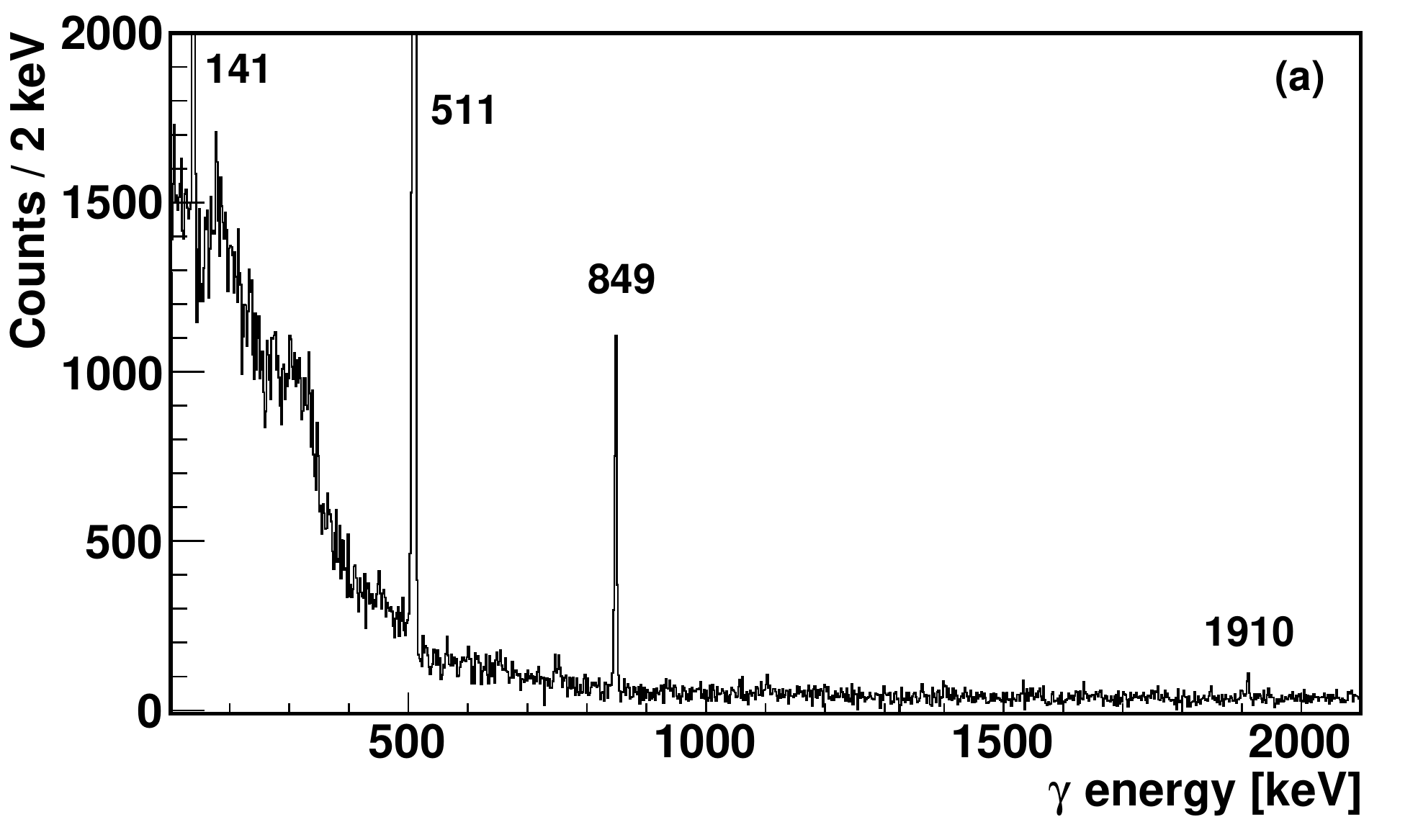}
	\end{minipage}
	\begin{minipage}{1.0\linewidth}
	  \includegraphics[width=1\columnwidth]{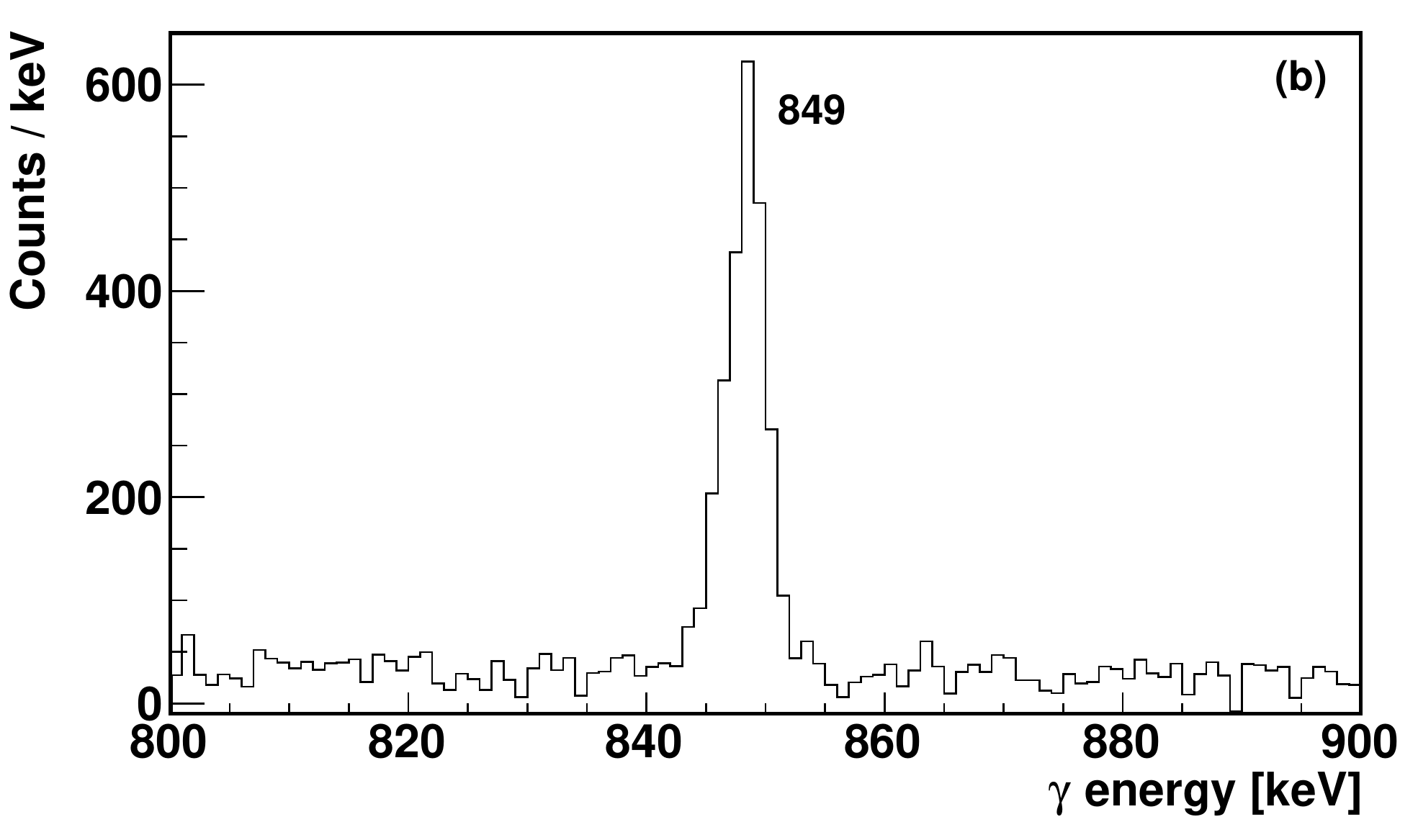}
	\end{minipage}
	\begin{minipage}{1.0\linewidth}
 	  \includegraphics[width=1\columnwidth]{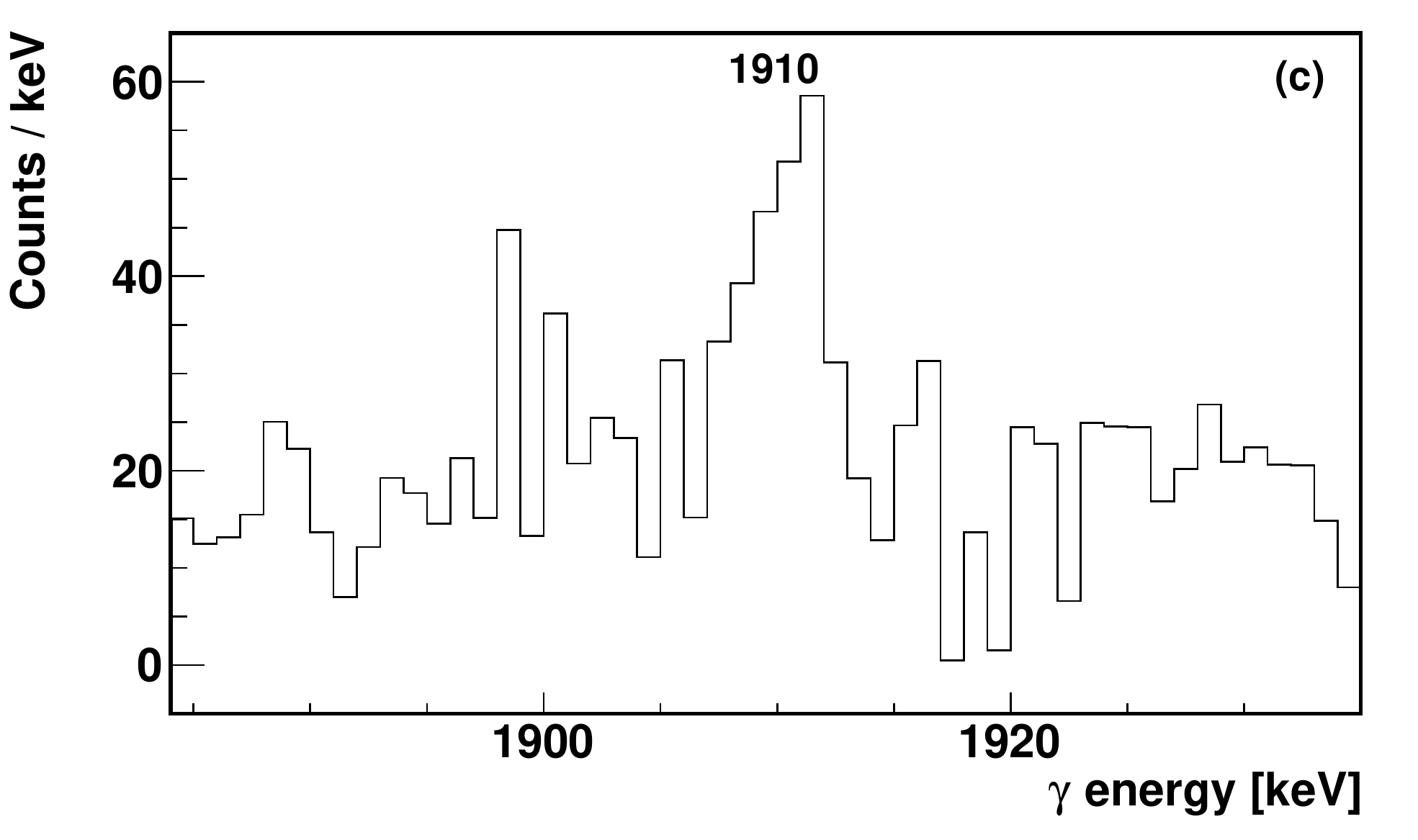}
	\end{minipage}
	\begin{minipage}{1.0\linewidth}
	  \includegraphics[width=1\columnwidth]{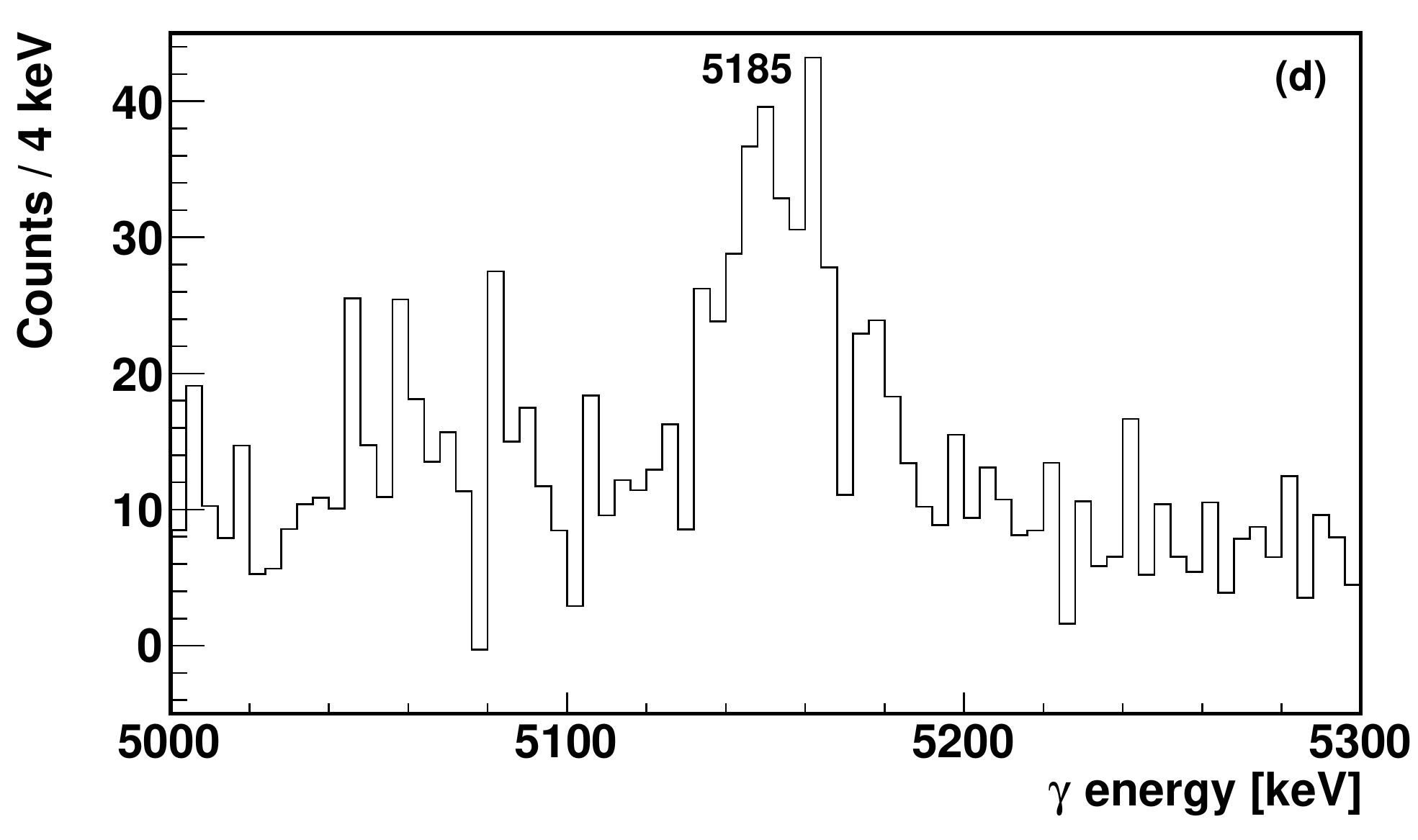}
	\end{minipage}
	\caption{(a) $\gamma$-ray spectrum observed for the decay of $^{52}$Ni. (b) Zoom of the 849 keV $\gamma$ line. (c) Zoom of the 1910 keV $\gamma$ line. (d) Zoom of the 5185 keV $\gamma$ line, detected using the low-amplification electronic chain (see text). The energy given for the peak includes the calibration made with the $\gamma$ lines from the decay of $^{52}$Co ($6^+$).}
	\label{gammas}
\end{figure}

\begin{figure}[!htb]
    \centering
		\vspace{-5.0 mm}
    \includegraphics[width=1\columnwidth]{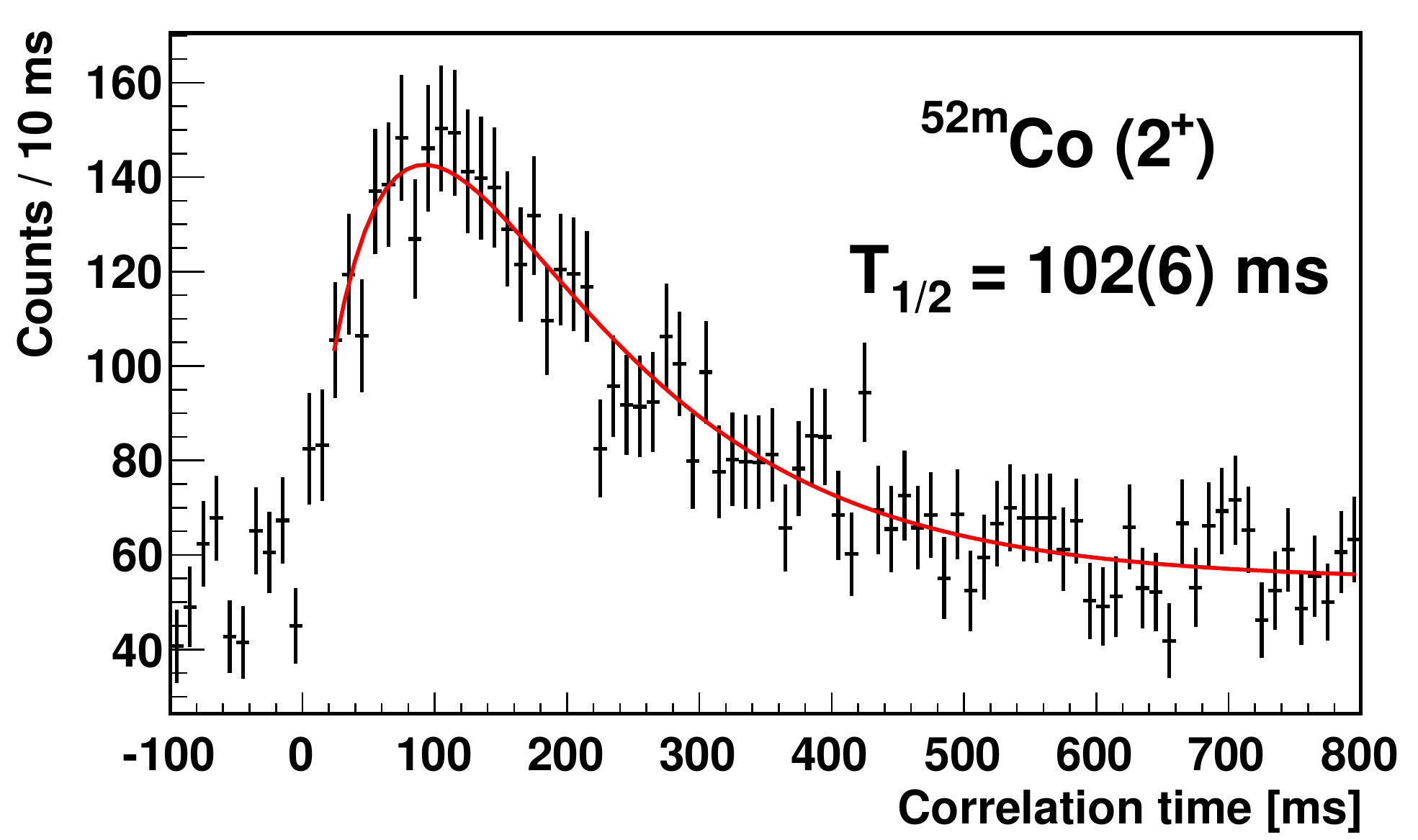}
	\caption{Fit of the correlation-time spectrum gated on the 849 keV $\gamma$ line, giving a $T_{1/2}$ = 102(6) ms for the $^{52}$Co, 2$^+$ isomer.}
	\label{halflife}
\end{figure}

To detect the population of the $^{52}$Co, 6$^+$ ground state would require the observation of four $\gamma$-rays in cascade (1329, 1942, 1535 and 849 keV) de-exciting the 6$^+$ IAS in $^{52}$Fe \cite{Hagberg1997183}, see Fig. \ref{decay}(b). In contrast the $\beta$ decay of the expected $^{52}$Co, 2$^+$ isomer should proceed to its IAS with 2 units of Fermi strength, and to a series of 1$^+$, 2$^+$ and 3$^+$ levels via GT transitions. Since there is no known 1$^+$ or 3$^+$ level below 6 MeV excitation energy in $^{52}$Fe, one can expect to observe the feeding to the IAS and some other 2$^+$ states. As discussed in Ref. \cite{Hagberg1997183}, the most intense $\gamma$-ray should be the 849 keV line \mbox{($2^+ \rightarrow 0^+$)}, which is also emitted in the cascade de-exciting the $6^+$ IAS in $^{52}$Fe. The specific signature of $^{52m}$Co (2$^+$) should be the strong population of the IAS, with the observation of its de-exciting $\gamma$-rays. In addition, in Ref. \cite{Hagberg1997183} it is proposed that the observation of a weak $\gamma$-ray at 1910 keV, belonging to a \mbox{$2^+ \rightarrow 2^+$} transition between the 2759 and 849 keV levels, should also be a typical feature of the population of $^{52m}$Co.

Ref. \cite{DECOWSKI1978186} reports two 2$^+$ levels in $^{52}$Fe separated in energy by 10 keV only, at 6034(5) and 6044(5) keV, both candidates to be the IAS of the $^{52}$Co, 2$^+$ isomeric state. If the existence of these 2$^+$ levels could be confirmed, they may provide another example of isospin mixing in the IAS. The mixing would be strong because of the very small energy separation. A similar situation has been observed, e.g., in Ref. \cite{PhysRevLett.112.222501} where the energy separation was of the order of 100 keV.

The $\gamma$-ray spectrum observed for the decay chain of $^{52}$Ni is shown in Fig. \ref{gammas}(a). In addition to the 141 keV $\gamma$-ray mentioned above (for the 2407 keV $\gamma$-ray see below) and the 511 keV $\gamma$ line associated with the annihilation of the positrons emitted in the $\beta$ decay, two other lines are observed at 849 and 1910 keV [they are also shown in Figs. \ref{gammas}(b) and (c), respectively]. The $\gamma$-ray seen at 1910 keV corresponds to a \mbox{$2^+ \rightarrow 2^+$} transition between the 2759 and 849 keV known levels \cite{NDS2015} in $^{52}$Fe [Fig. \ref{decay}(a)] and it is expected to be seen in the decay of the $^{52}$Co, 2$^+$ isomer \cite{Hagberg1997183}. The 1910 keV $\gamma$-ray, indeed, cannot be observed in the decay of the $^{52}$Co, 6$^+$ ground state because it does not populate the 2$^+$ state at 2759 keV. Moreover, the population of the 2759 keV state starting from the $^{52}$Fe, 4$^+$ level at 3584 keV would require the observation of a $\gamma$-ray of 825 keV, which we do not see [Fig. \ref{gammas}(b)].

The $\gamma$-ray spectrum shown in Fig. \ref{gammas} (a, b, c) was obtained using the high-amplification electronic chain \cite{PhysRevC.93.044336} and it allows the study of $\gamma$-rays up to 2 MeV. $\gamma$-rays of higher energy were detected using a low-amplification electronic chain, where a problem was observed during the data analysis consisting in a distortion of the peaks (see Ref. \cite{PhysRevC.93.044336} for details). In the $\gamma$-ray spectrum obtained with the low-amplification chain, in addition to the 2407 keV $\gamma$-ray in $^{52}$Co (good agreement was found in both energy and intensity when compared with values in the literature \cite{PhysRevC.93.044336,Dossat200718}), a $\gamma$-ray was observed at around 5 MeV [Fig. \ref{gammas}(d)]. The energy calibration at high energy was performed using the $\gamma$ lines observed in the decay of $^{52}$Co, $6^+$ (Section \ref{gs}). This procedure gave an energy of 5185(10) keV for the above $\gamma$-ray, which was then attributed to the \mbox{$2^+ \rightarrow 2^+$} transition between the $2^+$ IAS in $^{52}$Fe (at 6034(5) and/or 6044(5) keV \cite{DECOWSKI1978186,NDS2015}, where having one or both states does not change our conclusions) and the 849 keV level [Fig. \ref{decay}(a)]. A further confirmation comes from the fact that the 5185(10) keV $\gamma$-ray was also observed [Fig. \ref{gammasgs}(c) in Section \ref{gs}] when selecting events where $^{52}$Co was implanted, where one expects an admixture of both ground and isomeric states.

Therefore the observed 5185 keV $\gamma$-ray establishes clear evidence of the 2$^+$ isomer in $^{52}$Co, which is supported by the observed 1910 keV $\gamma$ line. Together, they constitute the first experimental evidence of the decay of $^{52m}$Co (2$^+$).

\begin{table}[!t]
  \vspace{-5.0 mm}
	\caption{$\gamma$-ray energies $E_{\gamma}$, $\gamma$ intensities $I_{\gamma}$ relative to $^{52}$Ni implants, and $\gamma$ intensities normalized to 100 decays from $^{52m}$Co (2$^+$) (using the intensity of the 141 keV $\gamma$-ray).}
	\label{table}
	\centering
	\begin{ruledtabular}
	  \begin{tabular}{r r r}
 		 $E_{\gamma}$ (keV) & $I_{\gamma}$ (\%) ($^{52}$Ni) & $I_{\gamma}/100$ decays (\%) ($^{52m}$Co) \\ \hline
			849(1)                          &        42(8)                 &                   97(26)                          \\
			1910(1)                        &         5(1)                  &                   12(3)                             \\
			5185(10)                      &        17(4)                 &                   39(12)                            \\
		\end{tabular}
	\end{ruledtabular}
\end{table}

\begin{table}[!t]
	\caption{Results on the $\beta^{+}$ decay of $^{52m}$Co (2$^+$) to $^{52}$Fe. Excitation energies $E_X$ in $^{52}$Fe, $\beta$ feedings $I_{\beta}$, Fermi $B$(F) and Gamow-Teller $B$(GT) transition strengths to the $^{52}$Fe levels.}
	\label{table2}
	\centering
	\begin{ruledtabular}
	  \begin{tabular}{c r r r}
		 $E_X$ (keV) & $I_\beta$ (\%)   & $B$(F)    & $B$(GT)   \\ \hline
			   849(1)        &       46(28)          &                  &   0.06(4)     \\
			 2759(2)        &      12(3)             &                  &   0.05(1)     \\
    6034(5)$^a$ - 6044(5)$^a$ & 39(12) & 1.6(5)$^b$ &                  \\
	  \end{tabular}
	\end{ruledtabular}
	\raggedright{$^a$ IAS, $E_X$ from Refs. \cite{DECOWSKI1978186,NDS2015}. \\ $^b$ Calculated assuming all the strength is Fermi.}
\end{table}

Besides the $\gamma$-rays described above, a $\gamma$-ray of 2760(1) keV was seen in Ref. \cite{JPSJ43} and attributed to a \mbox{$2^+ \rightarrow 0^+$} transition between the 2759 keV level and the ground state in $^{52}$Fe. Considering the intensity measured in Ref. \cite{JPSJ43} for this $\gamma$-ray and our low $\gamma$-efficiency at that energy, we do not expect to see this $\gamma$ line in our low-amplification spectrum, and indeed we do not observe it.

The half-life associated with a given $\gamma$ line is determined from the fit of the correlation-time spectrum gated on that $\gamma$ line, which was created according to the procedure described in Ref. \cite{PhysRevC.93.044336}. The fit performed for the 849 keV $\gamma$ line is shown in Fig. \ref{halflife}. The fit function includes the decay of the parent nucleus ($^{52}$Ni, with a known half-life of 42.8(3) ms \cite{PhysRevC.93.044336}, which was kept fixed), the growth of the daughter activity ($^{52m}$Co, of unknown half-life) and a constant background. From this fit we obtained a half-life of 102(6) ms for $^{52m}$Co (2$^+$). The much lower statistics prevented us from extracting the half-lives associated with the 1910 and 5185 keV $\gamma$-rays when selecting $^{52}$Ni implants. However, by selecting events where $^{52}$Co was implanted we were able to extract a half-life of 91(15) ms for the 5185 keV $\gamma$-ray, in agreement with the value quoted above.

The results are summarized in the decay scheme shown in Fig. \ref{decay}(a) and in Tables \ref{table} and \ref{table2}. The value \mbox{$Q_{\beta}$ = 11571(54) keV}, given in Fig. \ref{decay}(a) for the decay of $^{52}$Ni, was determined as explained in Section V of Ref. \cite{PhysRevC.93.044336}, where we deduced the ground state mass excesses for $^{52}$Ni and $^{52}$Co. Adding to that information the measured mass excess for $^{52}$Fe, -48332(7) keV \cite{Audi2012}, we can determine a value \mbox{$Q_{\beta}$ = 13845(52) keV} for the decay of the $^{52}$Co ground state, given in Fig. \ref{decay}(b).

Table \ref{table} gives the energies $E_{\gamma}$ and intensities $I_{\gamma}$ of the observed $\gamma$ peaks (both relative to $^{52}$Ni implants and normalized to 100 decays from $^{52m}$Co). The $\gamma$ intensities relative to $^{52}$Ni implants are determined as in Ref. \cite{PhysRevC.93.044336}. The $\gamma$-efficiency calibration, shown in Fig. 8 of Ref. \cite{PhysRevC.93.044336}, has been extended to higher $\gamma$ energies by Monte Carlo simulations. Since the $^{52}$Co $\gamma$-ray at 141 keV only populates $^{52m}$Co (2$^+$) and its intensity is 43(8)\%, the $\gamma$ intensities can be normalized to 100 decays from $^{52m}$Co using the intensity of the 141 keV $\gamma$-ray.

Table \ref{table2} gives the $\beta$ feedings $I_{\beta}$ and the Fermi and Gamow-Teller transition strengths for the $\beta$-decay of the $^{52}$Co, 2$^+$ isomer to $^{52}$Fe. The $\beta$ feedings to the levels populated in $^{52}$Fe are deduced from the $\gamma$ intensities for 100 decays from $^{52m}$Co (2$^+$). As discussed above, there are two possible candidate levels in $^{52}$Fe (at 6034(5) and 6044(5) keV \cite{DECOWSKI1978186,NDS2015}) for the IAS of $^{52m}$Co (2$^+$), which are expected to be strongly mixed.  Unfortunately the resolution of our low-amplification $\gamma$ spectrum did not allow the disentanglement of the two contributions based on the 5185 keV peak [Fig. \ref{gammas}(d)]. Moreover, in the population of the IAS, both Fermi and Gamow-Teller contributions are possible. Thus we have calculated an upper limit to $B$(F) assuming all the strength is due to the Fermi transition and taking an average excitation energy of 6039(7) keV. A maximum $B$(F) of 1.6(5) is obtained, in agreement with the expected value $|N-Z|$ = 2. This confirms that the intensity extracted for the 5185 keV $\gamma$-ray is meaningful, even with the peak distortion.

\section{\label{gs}Half-life of the $^{52}$Co, 6$^+$ ground state}

\begin{figure}[!b]
	\begin{minipage}{1.0\linewidth}
    \centering
    \includegraphics[width=1\columnwidth]{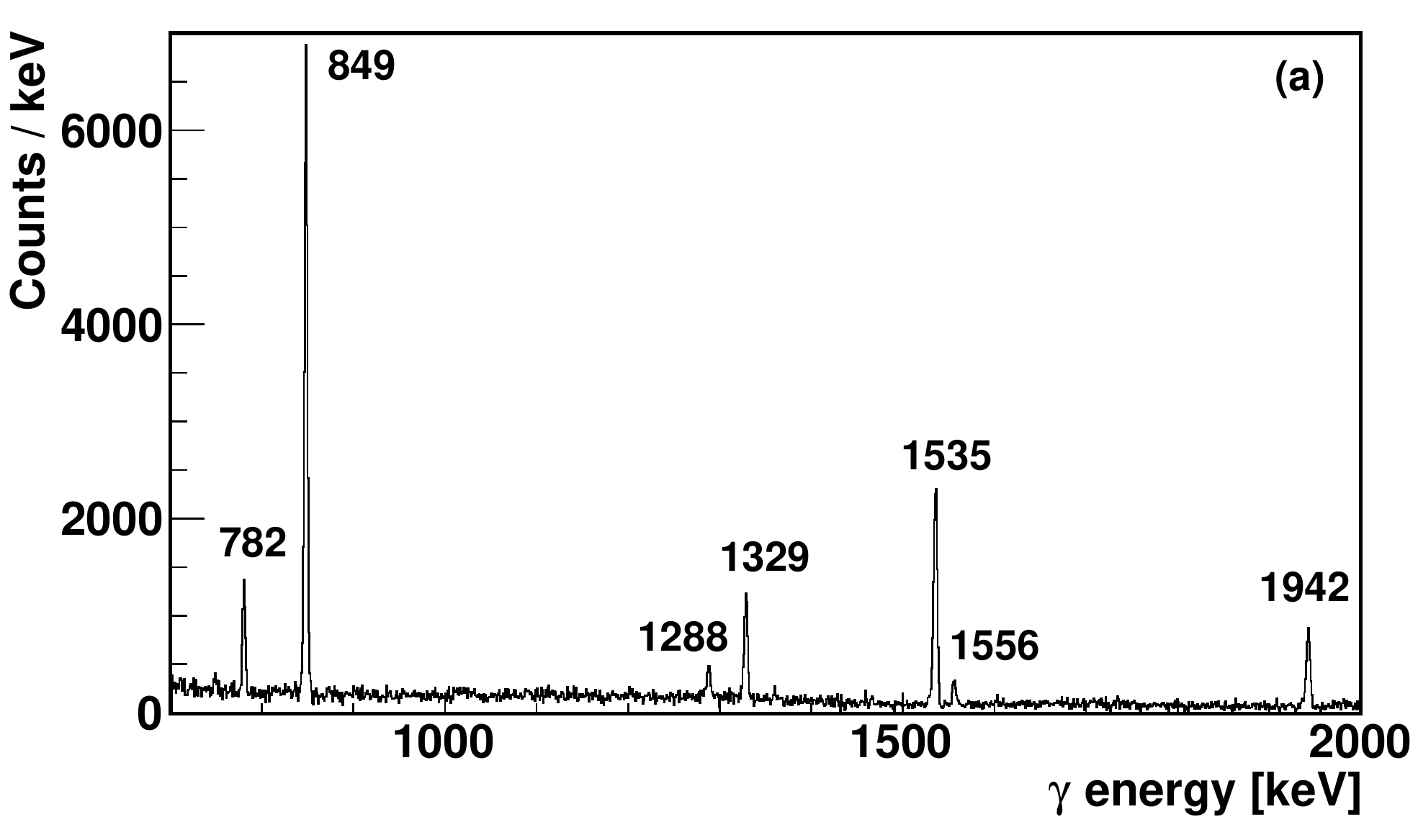}
	\end{minipage}
	\begin{minipage}{1.0\linewidth}
	  \includegraphics[width=1\columnwidth]{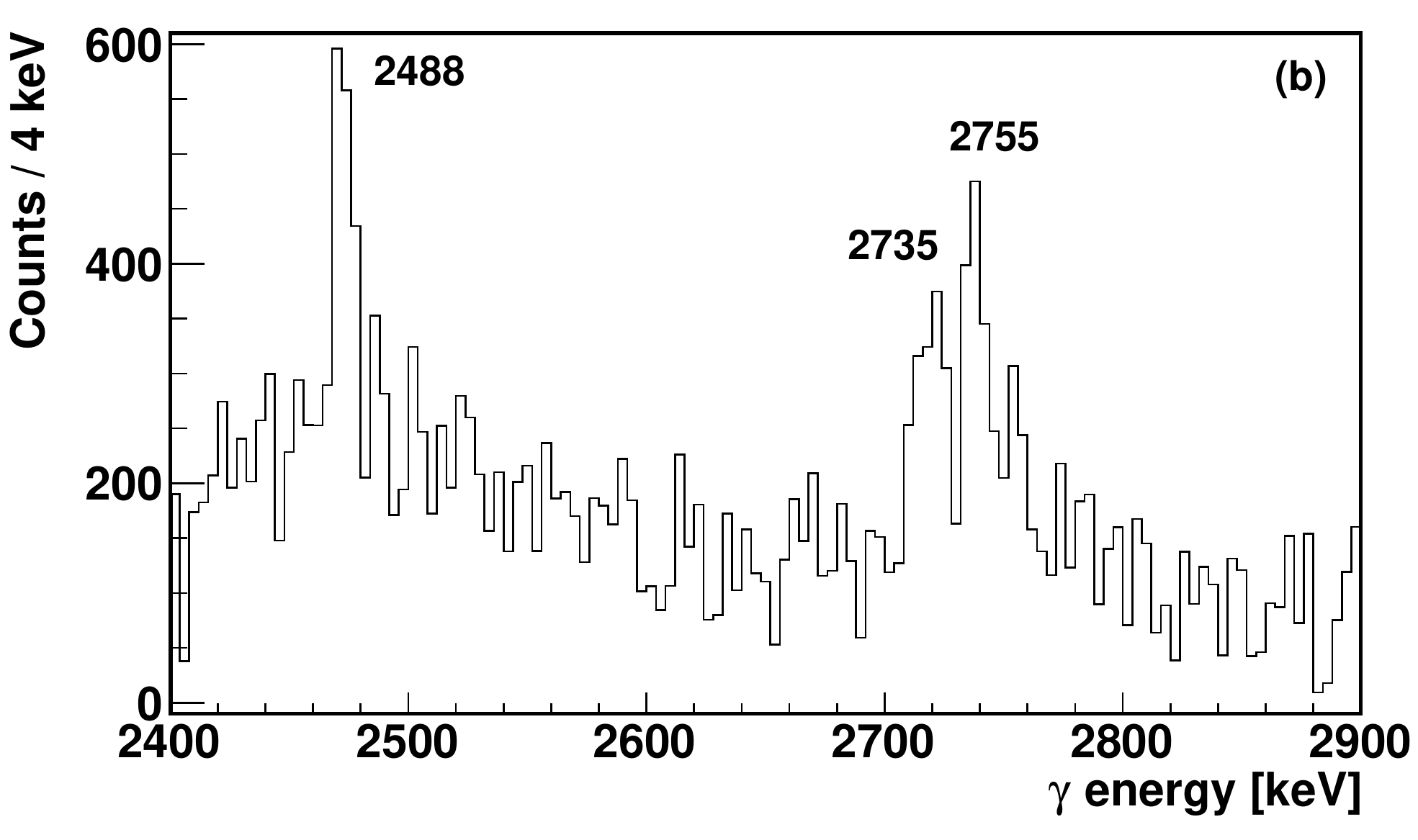}
	\end{minipage}
	\begin{minipage}{1.0\linewidth}
 	  \includegraphics[width=1\columnwidth]{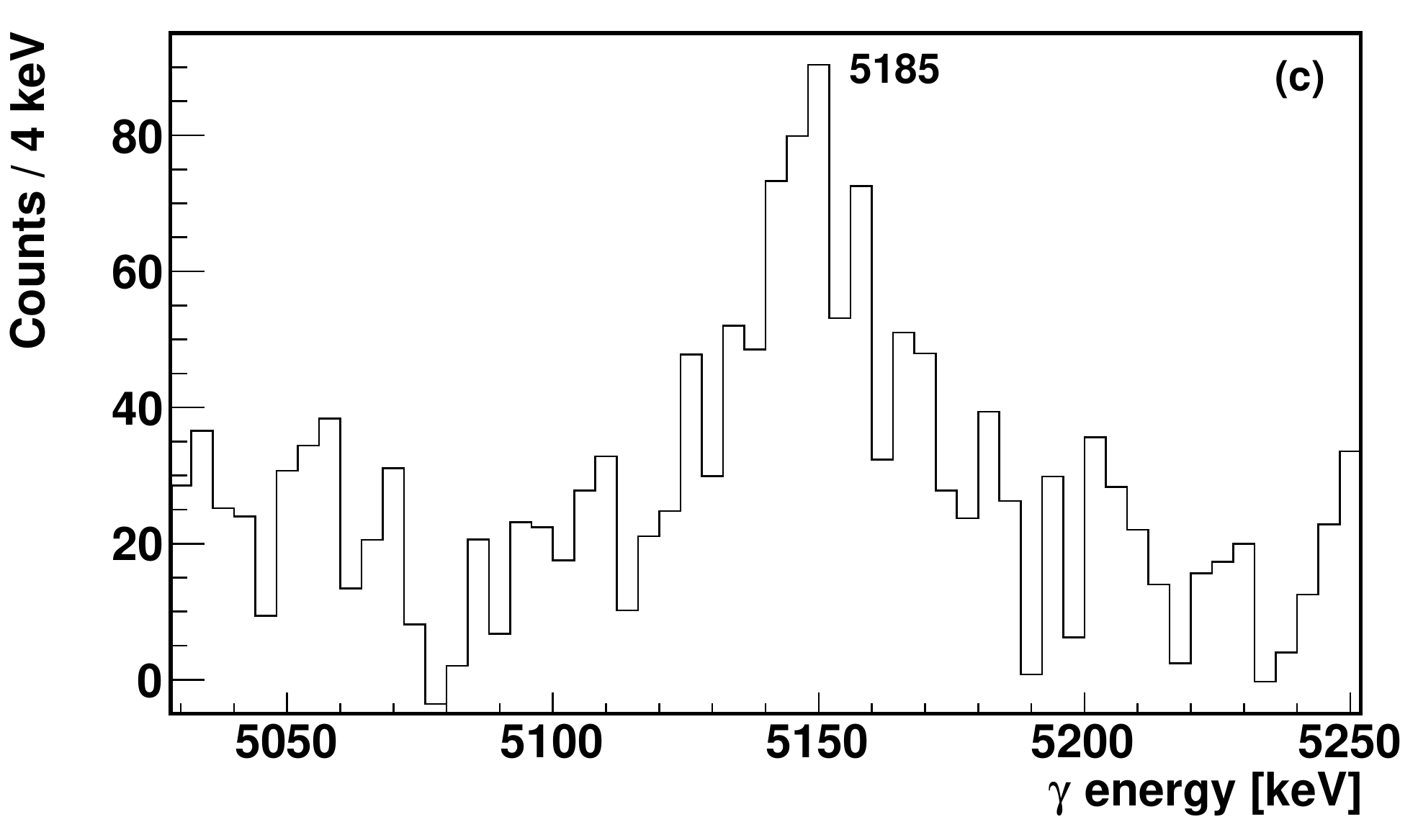}
	\end{minipage}
	\caption{(a) $\gamma$-ray spectrum observed for the decay of $^{52}$Co with the high-amplification electronic chain. (b) and (c) Zoom of the low-amplification spectrum in the regions of interest.}
	\label{gammasgs}
\end{figure}

In the same experiment $^{52}$Co fragments were also produced directly. This enabled us to add new information on the $\beta$ decay of the $^{52}$Co 6$^+$ ground state and measure its half-life with improved precision. In order to study the $\beta$ decay of the ground state, we have selected the events where $^{52}$Co was implanted (see Figs.  \ref{ID1plot} and \ref{ID2plot}). They should be a mixture of the ground and isomeric states. The high-amplification $\gamma$-ray spectrum obtained for the decay of $^{52}$Co is shown in Fig. \ref{gammasgs}(a). There, we observed known $\gamma$-rays (at 849, 1288, 1329, 1535, 1556 and 1942 keV) \cite{NDS2015}, expected from the decay of the levels populated in $^{52}$Fe, and a further $\gamma$-ray at 782 keV. In the low-amplification spectrum [Fig. \ref{gammasgs}(b)] we saw in addition other expected $\gamma$-rays from $^{52}$Fe (at 2488, 2735 and 2755 keV). In the latter spectrum we also observed the 5185(10) keV $\gamma$-ray from the decay of $^{52m}$Co (2$^+$) [Fig. \ref{gammasgs}(c)].

The $\beta$ decay of the $^{52}$Co, 6$^+$ ground state is summarized in Fig. \ref{decay}(b), where $\beta$ feeding is expected to the 5$^+$, 6$^+$ and 7$^+$ levels in $^{52}$Fe. The 6$^+$ IAS at 5656 keV \cite{NDS2015} in $^{52}$Fe de-excites by $\gamma$-ray cascades starting with the 782 and 1329 keV $\gamma$-rays. A possible 516 keV $\gamma$-ray connecting the IAS and the 5140 keV level would be hidden below the 511 keV annihilation peak.

\begin{figure}[!th]
    \centering
		\includegraphics[width=1\columnwidth]{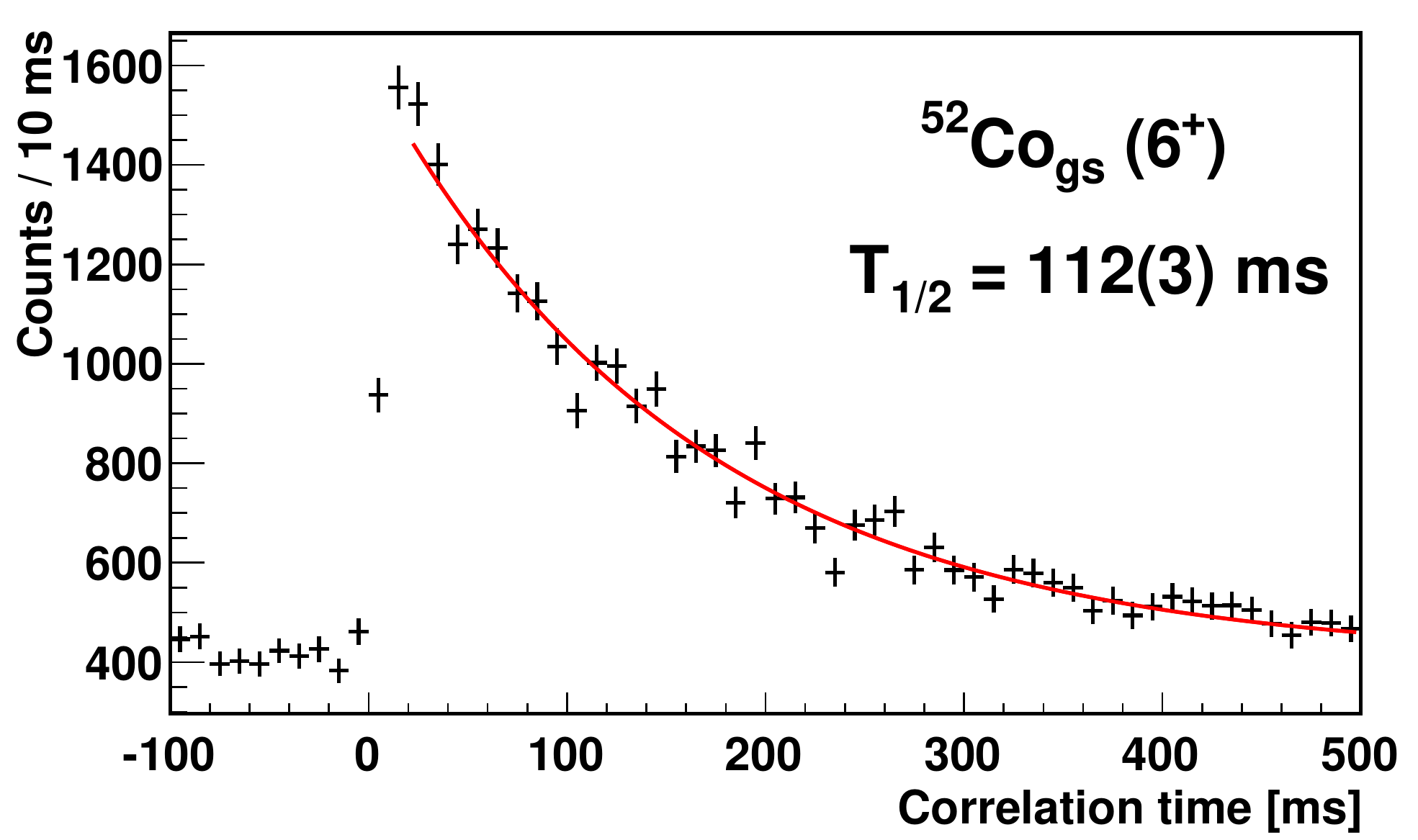}
	\caption{Fit of the correlation-time spectrum obtained as a sum of the spectra gated on the 1329, 1535 and 1942 keV $\gamma$-rays from the decay of the $^{52}$Co, 6$^+$ ground state, giving a $T_{1/2}$ = 112(3) ms.}
	\label{halflifegs}
\end{figure}

We have observed two $\gamma$-rays at 1556 and 2755 keV, corresponding to the de-excitation of the level at 5140(5) keV. A level at 5134(8) keV was observed in Ref. \cite{DECOWSKI1978186}, where a \mbox{$J^{\pi}$ = 5$^-$} was attributed to it. In Ref. \cite{JPSJ43} a level was seen at 5138(4) keV, de-exciting by two $\gamma$-rays of 2380 and 4286 keV which we do not observe. Finally, in Ref. \cite{PhysRevC.58.3163} a level was observed at 5137 keV, de-exciting by three $\gamma$-rays of 740.6, 1553 and 2753 keV. The last two $\gamma$-ray energies agree marginally with our observed $\gamma$-rays at 1556 and 2755 keV, however we did not observe the 740.6 keV $\gamma$-ray, which is supposed to be stronger than the 1553 keV $\gamma$ line according to Ref. \cite{PhysRevC.58.3163}. We believe we see the same level as in Ref. \cite{PhysRevC.58.3163}, and not the level reported in Ref. \cite{JPSJ43}. We also do not know if the level we observed corresponds to that in Ref. \cite{DECOWSKI1978186}, consequently we have put the 5$^-$ assignment in parenthesis in Fig. \ref{decay}(b).

Table \ref{table3} gives the energies $E_{\gamma}$ of the observed $\gamma$ peaks (first column) and their intensities $I_{\gamma}$ normalized to the 849 keV $\gamma$-ray (second column). We could not extract the intensities for the $\gamma$-rays above 2 MeV, only observed in the low-amplification spectrum affected by the peak distortion. We obtained the intensity of the 2735 keV $\gamma$-ray by summing those of the 1288 and 1556 keV $\gamma$-rays, i.e., we assumed that the 3584 keV level is not directly populated in either the $\beta$ decay of the 6$^+$ ground state or the $\beta$ decay of the 2$^+$ isomeric level.

Looking at the intensities normalized to that of the 849 keV $\gamma$-ray, we get 24(6)\% and 27(7)\% for the 1329 and 1942 keV $\gamma$-rays, respectively. Within the errors, a small amount of $\beta$ feeding to the 4327 keV level is possible. In Ref. \cite{Hagberg1997183} the intensity of the 1942 keV $\gamma$-ray was reported to be 17\% lower than that of the 1329 keV $\gamma$-ray. This was probably because the 1942 keV peak was not resolved from a 1944 keV peak from the decay of $^{50m}$Mn.

The summed intensities of the 1535 and 2735 keV $\gamma$-rays, which both go to the 849 keV level, are 83(31)\%. This means that the 2$^+$ level at 849 keV may have an extra feeding of 17(31)\% that may be attributed to the $\beta$ decay of the $^{52}$Co (2$^+$) isomeric state. With this information one can normalize the $\gamma$ intensities to 100 decays from the $^{52}$Co (6$^+$) ground state; these values are presented in the third column of Table \ref{table3}. They are also used to calculate the $\beta$ feedings shown in the second column of Table \ref{table4}. We expect that the levels at 4872 and 5140 keV get some direct feeding, which we cannot estimate because we miss the intensities of the 2488 and 2755 keV $\gamma$-rays. Thus we attribute the missing $\beta$ feeding (50\%) to these 4872 and 5140 keV levels. Besides the $\beta$ feedings, Table \ref{table4} gives $B$(F) and $B$(GT). Also in this case both Fermi and Gamow-Teller contributions are possible in the population of the IAS, thus we calculated an upper limit to $B$(F). We obtained a maximum $B$(F) of 1.7(3), which agrees with the expected value $|N-Z|$ = 2.

\begin{table}[!t]
  \vspace{-5.0 mm}
	\caption{Column one shows the $\gamma$-ray energies $E_{\gamma}$. Column two gives the $\gamma$ intensities $I_{\gamma}$ relative to the 849 keV $\gamma$-ray, including both the 6$^+$ and the 2$^+$ $^{52}$Co decays. Column three presents the $\gamma$ intensities normalized to 100 decays from the $^{52}$Co (6$^+$) ground state.}
	\label{table3}
	\centering
	\begin{ruledtabular}
	  \begin{tabular}{r r r}
 		 $E_{\gamma}$ (keV) & $I_{\gamma}$/$I_{\gamma}$(849) (\%) & $I_{\gamma}/100$ decays (\%) ($^{52}$Co$_{gs}$) \\ \hline
		  782(1)                          &        15(4)                    &                   18(5)                        \\
      849(1)                          &      100(26)                 &                 100(21)                       \\
    1288(1)                          &          8(2)                    &                   10(3)                         \\
    1329(1)                          &        24(6)                    &                   29(8)                         \\
    1535(1)                          &        67(17)                  &                   81(21)                       \\
    1556(1)                          &          7(2)                    &                     9(2)                          \\
    1942(1)                          &        27(7)                    &                   32(8)                          \\
    2488(5)                          &                                      &                                                      \\
    2735(5)                          &        16(3)                    &                   19(3)                           \\
    2755(5)                          &                                      &                                                       \\
		\end{tabular}
	\end{ruledtabular}
\end{table}

\begin{table}[!t]
	\caption{Results on the $\beta^{+}$ decay of the $^{52}$Co (6$^+$) ground state to $^{52}$Fe. Excitation energies $E_X$ in $^{52}$Fe, $\beta$ feedings $I_{\beta}$, Fermi $B$(F) and Gamow-Teller $B$(GT) transition strengths to the $^{52}$Fe levels.}
	\label{table4}
	\centering
	\begin{ruledtabular}
	  \begin{tabular}{c r r r}
		 $E_X$ (keV) & $I_\beta$ (\%)   & $B$(F)    & $B$(GT)   \\ \hline
		  849(1)           &                            &                  &                    \\
    2385(1)           &                            &                  &                    \\
    3584(5)           &                            &                  &                    \\
    4327 (2)           &      3(11)            &                  &   0.03(12)  \\
    4872(5)          & \rdelim\}{2}{8mm}[$\leq$50] &  &           \\
    5140(5)           &                           &                   &                    \\
    5656(2)$^a$   &       47(9)          & 1.7(3)$^b$ &                    \\
	  \end{tabular}
	\end{ruledtabular}
	\raggedright{$^a$ IAS. \\ $^b$ Calculated assuming all the strength is Fermi.}
\end{table}

As mentioned above, in fragmentation experiments both the 6$^+$ ground state and the 2$^+$ isomer will be implanted together and cannot be separated with the available information on the implants. This has to be taken into account in the determination of the half-life of the $^{52}$Co ground state. In Ref. \cite{Hagberg1997183}, indeed, because of the ambiguity of the origin of the 849 keV $\gamma$-rays their apparent half-life [104(11) ms] was not used to determine the half-life of the $^{52}$Co ground state [$T_{1/2}$ = 115(23) ms]. More recently, the $\beta$ decay of $^{52}$Co was revisited in Ref. \cite{PhysRevC.87.024312} and a value of 103(7) ms was extracted for the half-life of the ground state by gating on the $\beta$ particles. Combining this with the previous measurement \cite{Hagberg1997183} gives a weighted average value of 104(7) ms which is the value reported in the most recent compilation for mass $A$ = 52 \cite{NDS2015}. However, in Ref. \cite{PhysRevC.87.024312} the possible implantation of the $^{52}$Co isomer together with the $^{52}$Co ground state was not considered.

To determine the half-life of the $^{52}$Co, 6$^+$ ground state in a isomer-free way, we have constructed a correlation-time spectrum as the sum of the spectra gated on the 1329, 1535 and 1942 keV $\gamma$-rays. The fit to this spectrum, shown in Fig. \ref{halflifegs}, gives $T_{1/2}$ = 112(3) ms. This result agrees with the value from Ref. \cite{Hagberg1997183} but the precision is improved.

\section{\label{concl}Conclusions}

We reported the first experimental observation of the decay of the 2$^+$ isomeric state in $^{52}$Co, which was produced in the $\beta$ decay of $^{52}$Ni. We observed the decay of $^{52m}$Co to $^{52}$Fe, where it populates various 2$^+$ states including the IAS. These 2$^+$ levels then de-excite by $\gamma$-ray emission and we observed three $\gamma$-rays at 849, 1910 and 5185 keV. The observed de-excitation of the IAS (by the 5185 keV $\gamma$-ray) is clear evidence of the population of the 2$^+$ isomer, which is reinforced by the observation of the expected \cite{Hagberg1997183} $\gamma$-ray at 1910 keV. The $\beta$ feedings for the decay of the $^{52}$Co isomer to the 2$^+$ levels in $^{52}$Fe and the Fermi and Gamow-Teller transition strengths have been determined. We have extracted a half-life of 102(6) ms for the $^{52}$Co, 2$^+$ isomer using the 849 keV $\gamma$ line.

We have also studied the $\beta$ decay of the $^{52}$Co, 6$^+$ ground state by gating on the events where $^{52}$Co was implanted, obtaining new information. Many $\gamma$-rays were observed, including a previously unobserved $\gamma$-ray at 782 keV, and their intensities were determined. The $\beta$ feedings for the decay of $^{52}$Co (6$^+$) to the 6$^+$ levels in $^{52}$Fe and the $B$(F) and $B$(GT) were deduced. A half-life of 112(3) ms was measured for the $^{52}$Co (6$^+$) ground state, improving the uncertainty in comparison with the values reported in the literature.

The $^{52}$Co nucleus lies in the $rp$-process pathway, where the proton-absorption and $\beta$-decay processes compete. Hence the existence of a $\beta$-decaying isomer as well as its decay properties are important.

\begin{acknowledgments}
This work was supported by the Spanish MICINN grants FPA2008-06419-C02-01, FPA2011-24553, FPA2014-52823-C2-1-P; Centro de Excelencia Severo Ochoa del IFIC SEV-2014-0398; $Junta~para~la~Ampliaci\acute{o}n~de~Estudios$ Programme (CSIC JAE-Doc contract) co-financed by FSE; ENSAR project 262010; MEXT, Japan 18540270 and 22540310; Japan-Spain coll. program of JSPS and CSIC; UK Science and Technology Facilities Council (STFC) Grant No. ST/F012012/1; Region of Aquitaine. E.G. acknowledges support by TUBITAK 2219 International Post Doctoral Research Fellowship Programme. R.B.C. acknowledges support by the Alexander von Humboldt foundation and the Max-Planck-Partner Group. We acknowledge the EXOGAM collaboration for the use of their clover detectors.
\end{acknowledgments}

\bibliography{references}

\end{document}